\documentclass[11pt]{article}
\pdfoutput=1
\usepackage{graphicx}
\usepackage{amsmath}
\usepackage{amssymb}
\usepackage{slashed}
\usepackage{epsfig}
\usepackage{hyperref}
\usepackage{amsbsy}
\usepackage{textcomp}
\usepackage{geometry}
\newcommand{\eps}{\epsilon}
\newcommand{\nn}{\nonumber}
\newcommand{\T}{\rule{0pt}{3.0ex}}
\newcommand{\B}{\rule[-2.2ex]{0pt}{0pt}}

\begin{document}

\title{\bf{A Note on the IR Finiteness of Fermion Loop Diagrams \\}}
\author{
{Ambresh Shivaji \footnote{email:ambreshkshivaji@hri.res.in}} \\
{Harish-Chandra Research Institute}, \\ {Chhatnag Road, Junsi}, \\ {Allahabad-211019, India} 
       }

\date{}

\maketitle{}

\begin{center}
 {\bf Abstract}
\end{center}

 { \noindent We show that the most general fermion loop diagram is finite in both soft and collinear regions and therefore, 
  it's IR finite. We use this result to express the IR singular structure of a box scalar integral in terms of 
  the IR singular structure of reduced triangle scalar integrals.}


\section{Introduction}

The structure of infrared singularities of Feynman graphs is very well understood at one-loop level \cite{Kinoshita:1962ur, Collines89}. 
These infrared singularities appear due to the massless particles present in the theory. We may therefore call them 
``mass singularities'' as well. In literature, these mass singularities are classified as soft and collinear singularities. 
The appearance of soft singularity in loop diagrams is associated with the exchange of massless particles between two on-shell 
particles. On the other hand, splitting of a massless external particle into two massless internal particles of a loop diagram 
gives rise to collinear singularity. In fact, these are only the necessary conditions for a loop diagram to have soft or collinear 
divergence. Actual singularities appear as any of the momenta of internal lines, in these configurations, vanishes (soft divergence)
or it becomes parallel to one of its neighboring external legs (collinear divergence). Both of these singularities are logarithmic 
in $n=4$ dimensions.

A soft singular configuration of loop diagrams, with massless external as well as internal particles, contains two collinear
configurations and this situation leads to the possibility of overlapping of soft and collinear singularities \footnote{Note 
that at one-loop, overlapping of two soft regions or two distant collinear regions does not take place in general.}. The 
overlapping singularity is also logarithmic in nature. In dimensional regularization ($ n=4-2\eps_{IR}, \eps_{IR} \rightarrow 0^-$), 
IR divergence of a loop integral appears as $1/\eps_{IR}$ poles. The IR divergent scalar one-loop integrals have generic form given 
by $ \sim  \frac{A}{{\eps_{IR}}^2} + \frac{B}{\eps_{IR}} + C + \mathcal{O} (\eps_{IR}) $, where $A, B, C$ are complex functions of 
kinematic invariants \cite{Ellis:2007qk}. The $1/{\eps_{IR}}^2 $ term corresponds to the overlapping of soft and collinear divergence.
It should be obvious that unlike in the case of UV singularity, tensor integrals do not spoil the (logarithmic) structure of IR 
divergence of scalar integrals. In Ref.~\cite{Dittmaier:2003bc} an expression to determine $A$ and $B$ for a general $N$-point 
one-loop tensor integral is obtained. A very naive way of understanding these mass singularities at one-loop can be found 
in~\cite{Shivaji:2010re}. 

There are many scattering processes which, at one-loop, proceed via fermion loop diagrams. In fact some of these scattering 
processes are possible only at one-loop {\it i.e.} there is no tree level diagram for them at the leading order. Such 
processes at one-loop, are UV as well as IR finite and this is normally expected when contribution from all the relevant 
one-loop diagrams are included. A well known standard example is of four photon interaction in QED.  Although these fermions 
in the loop are not exactly massless, at very high energies their masses can be neglected. As discussed above, working in 
vanishing fermion mass limit may lead to mass singularities. In this case, the singularities would show up as large-logs of 
vanishing fermion masses. This is another way of regularizing IR singularities, often called ``mass regularization'' in literature.
In mass regularization, the generic form of fermion loop diagrams is, $\sim A \; {\rm ln}^2(m^2) + B \; {\rm ln}(m^2) + C + \mathcal{O}(m^2)$ 
\footnote{ In general there is no one-to-one correspondence between results obtained using dimensional regularization and mass 
regularization.}. In this form, ${\rm ln}^2(m^2)$ piece refers to the overlapping singularity.\\
 
In the following we show that for a given fermion loop diagram, $ A=B=0$, that is, it is IR finite. The proof is obvious for 
the case of massive fermions in the loop. Also, if all the external legs are off-shell, the diagram would be IR finite even 
for the case of massless fermions in the loop. So we need to consider only those fermion loop diagrams in which fermions are 
massless or more correctly their masses can be neglected and at least one external leg is massless. The main result of this 
paper is partially contained in Ref.~\cite{Nagy:2003qn}. Although the result may be known to some experts in the field, we 
aim at making this interesting result more accessible to the community.   


\section{Proof of the Main Result }

\begin{figure}[h!]
\begin{center}
\includegraphics[width=5cm]{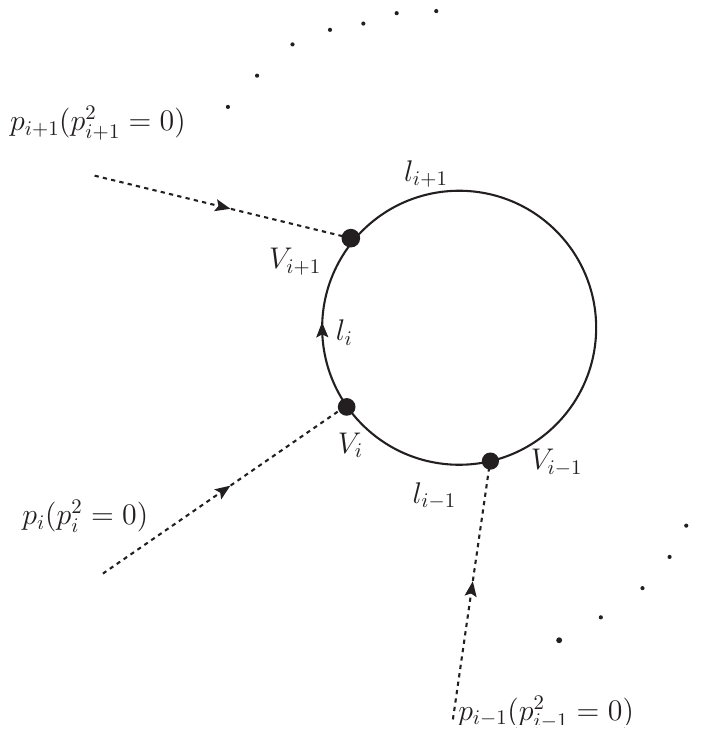}
\caption{ Massless fermion loop diagram. Dotted external lines represent any suitable massless particle. }
\label{fig:fermionloop}
\end{center}
\end{figure}

\begin{figure}[h]
\begin{center}
\includegraphics[width=11cm]{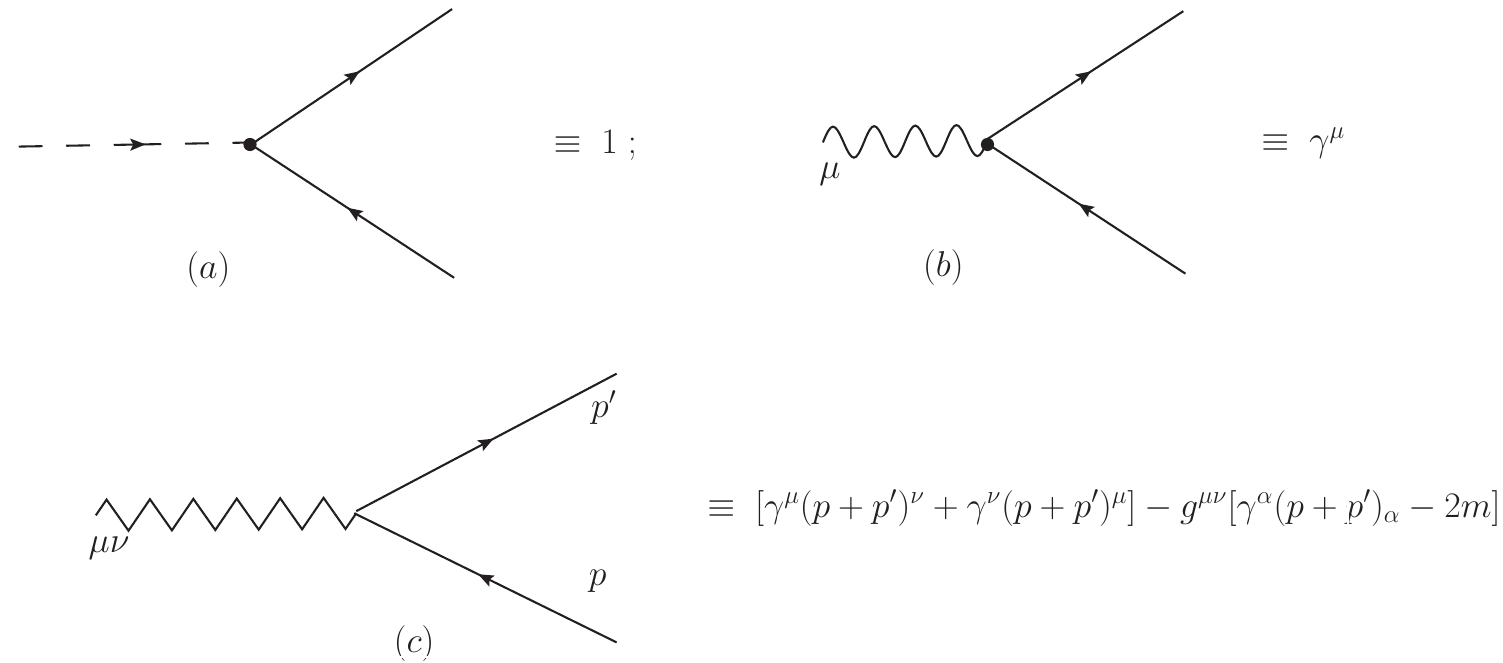}
\caption{Feynman rules: (a) scalar-fermion-fermion vertex 
(b) vector boson-fermion-fermion vertex (c) graviton-fermion-fermion vertex }
\label{fig:feynrules}
\end{center}
\end{figure}

In general, we can have scalars, gauge bosons and gravitons as external particles attached to a fermion loop. With one massless 
external particle we expect only collinear singularity while for two adjacent external massless particles, soft and collinear 
and their overlap may develop. IR finiteness of a fermion loop diagram can be shown by showing its soft finiteness and collinear 
finiteness. This automatically takes care of its finiteness in overlapping regions. The general fermion loop integral has following 
form ( see Fig.~\ref{fig:fermionloop}), 
 
\begin{equation}\label{eq:floop_int}
 I \simeq \int d^nl\; {{\rm tr}(....\slashed l_{i-1}\;V_i\; \slashed l_i \; V_{i+1}\; \slashed l_{i+1}....)\over
....l_{i-1}^2\;l_i^2\;l_{i+1}^2....},
 \end{equation}
where $V_i$s are vertex factors for a given massless external particle attached to the fermion loop. From momentum conservation at 
each vertex it is clear that $l_i = l_{i-1} + p_i, l_{i+1} = l_i + p_{i+1}$ etc. The Feynman rules for vertices of interest are 
given in Fig.~\ref{fig:feynrules}.  Factors of `$i$' and coupling constants etc. are dropped in writing those rules, for simplification.
Feynman rules for graviton-fermion interaction is derived in \cite{Holstein06}. We did not consider two graviton-fermion vertex since 
it does not correspond to the soft or collinear configuration described above. We closely follow the analysis of IR singular structure 
of one-loop diagrams, given in Ref.~\cite{Shivaji:2010re}.\\


First we would like to see the behavior of this integral as any of the internal lines becomes soft {\it i.e.} its momentum vanishes.  
Without loss of generality we consider softness of $l_i$ and take $l_i = \eps$. We see that in the limit $\eps \rightarrow 0$, the 
denominators which vanish in general are,
\begin{eqnarray}\label{eq:soft_deno}
 l_{i-1}^2 &=& \eps^2 - 2\eps \cdot p_i, \nonumber \\
l_i^2 &=& \eps^2  \nonumber \\
{\rm and}\;\; l_{i+1}^2 &=& \eps^2 + 2\eps \cdot p_{i+1},
\end{eqnarray}
where we have used on-mass-shell conditions for $p_i$ and $p_{i+1}$. Neglecting $\eps^2$ with respect to $\eps\cdot p_i$ and 
$\eps\cdot p_{i+1}$ in Eq.~\ref{eq:soft_deno}, we see that the integral in Eq.~\ref{eq:floop_int}, in the soft limit, behaves as
\begin{equation}
 I \sim \int d^n\eps\; \frac{\slashed \eps}{\eps\cdot p_i\; \eps^2\; \eps\cdot p_{i+1}} \sim \eps^{n-3}
\end{equation}
and it vanishes in $n=4$ dimensions. Thus each fermion loop diagram is soft finite, independent of kind of massless external 
particles attached to it. We should mention here that the soft finiteness of fermion loop diagrams is indirectly shown by 
Kinoshita in \cite{Kinoshita:1962ur}. \\


Next we consider the behavior of fermion loop integral in collinear region. We take {$ l_i = x p_{i+1}+\epsilon_\perp$}, 
($x\ne 0,-1$ as it corresponds to softness of $l_i$ and $l_{i+1}$ respectively) with $\eps_\perp\cdot p_{i+1} = 0 $. 
Note that in $\eps_\perp \rightarrow 0$ limit, this condition implies collinearity of $l_i$ with $p_{i+1}$. In this 
collinear limit only vanishing denominators are $l_i^2 = \eps_\perp^2$ and $l_{i+1}^2 = \eps_\perp^2$. Thus the 
integral in Eq.~\ref{eq:floop_int} reads,
\begin{equation}
 I \simeq \int {d^n\epsilon_\perp \over\epsilon_\perp^4}\; {\rm tr} (...\slashed l_{i-1} \; V_i \; 
 \slashed p_{i+1} \; V_{i+1} \; \slashed p_{i+1}... ).
\end{equation}
We need to make the above substitution for $l_i$ in $V_i$s also, in case they depend on the loop momentum {\it e.g.} in 
graviton-fermion vertex. Since the vertex factors are different for different kind of external particles 
(see Fig.~\ref{fig:feynrules}), we will consider three separate cases to see the behavior of fermion loop 
integral in the collinear limit. 
\begin{enumerate}
 \item 
{\bf Scalars :} 

In this case, the vertex factor is simply $V_{i+1} = 1 $, and since $\slashed p_{i+1}\slashed p_{i+1} = p_{i+1}^2 = 0$,
the integral in Eq.~(4) is
\begin{eqnarray}
 I \simeq \int {d^n\epsilon_\perp \over\epsilon_\perp^4} {\rm tr} (...\slashed l_{i-1}\slashed p_{i+1}\slashed p_{i+1}... ) = 0.
\end{eqnarray}

\item
{\bf Vector bosons :} 

The vertex factor in this case is, 
\begin{equation}
 V_{i+1} = \gamma_\mu e_{i+1}^\mu = \slashed e_{i+1} .
\end{equation}
 Here $e_{i+1}^\mu$ is the polarization vector of the gauge boson with momentum $p_{i+1}$. 
 Using transversality and on-shell condition for vector boson, we see that 
\begin{eqnarray}
 \slashed p_{i+1}\slashed e_{i+1}\slashed p_{i+1} &=& 2 \slashed p_{i+1} e_{i+1}\cdot p_{i+1}-p_{i+1}^2 \slashed e_{i+1} \nn \\
                                                  &=&   0,
\end{eqnarray}
and therefore the fermion loop integral in Eq.~(4) vanishes in the collinear limit.
\item
{\bf Gravitons :} 

The graviton-fermion vertex factor is given by
\begin{eqnarray}
 V_{i+1} =&& \Big[\gamma_\mu(2l_i+p_{i+1})_\nu + \gamma_\nu(2l_i+p_{i+1})_\mu \nonumber \\
          && - g_{\mu\nu}(2\slashed l_i+\slashed p_{i+1}-2m)\Big] e_{i+1}^{\mu\nu},
\end{eqnarray}
where $e_{i+1}^{\mu\nu}$ is polarization tensor for graviton of momentum $p_{i+1}$. It has following 
well known properties,

\begin{eqnarray}
 e_{i+1}^{\mu\nu}g_{\mu\nu} = ({e_{i+1}})^\mu_\mu &=& 0 \; \mbox{(traceless condition)}, \nonumber\\
p_{i+1}^\mu ({e_{i+1}})_{\mu\nu} &=& 0 \; \mbox{(transverse condition)}.
\end{eqnarray}

Using these properties the vertex factor in Eq.~(8) becomes
\begin{equation}
 V_{i+1} = 4 \gamma_\mu e_{i+1}^{\mu\nu}(l_i)_\nu.
\end{equation}
In the collinear limit, taken above, it is
\begin{equation}
 V_{i+1} = 4 x \gamma_\mu e_{i+1}^{\mu\nu}(p_{i+1})_\nu = 0,
\end{equation}
due to the transverse condition. Therefore the fermion loop diagram with external gravitons, like the cases
of scalars and vector bosons, is also collinear finite. 
\end{enumerate}

Combining all the above results of this section we can conclude that any fermion loop diagram 
of practical interest is always IR finite. The result holds even for axial coupling of external particles 
with the fermion. Although we have not considered any flavor change in the loop, it should be clear from 
the above analysis that our result remains true for any possible flavor changing interaction vertex in the loop. 

\section{An Application}

We can utilize the above fact regarding an individual fermion loop diagram, to show 
that the IR structure of any one-loop amplitude with $N\ge 3$, can be fixed completely in terms
of the IR structure of three-point (triangle) functions only~\cite{Dittmaier:2003bc}. Since, in 4 
dimensions, any $N$-point one-loop amplitude can be written in terms of tadpole, bubble, triangle 
and box scalar integrals, it would be sufficient for us to show that the IR singularities
of box scalar integrals are expressible in terms of those of reduced triangle scalars. A reduced 
triangle scalar is obtained by removing one of the four denominators in a given box scalar integral.
\begin{figure}[h]
\begin{center}
\includegraphics [angle=0,width=0.4\linewidth] {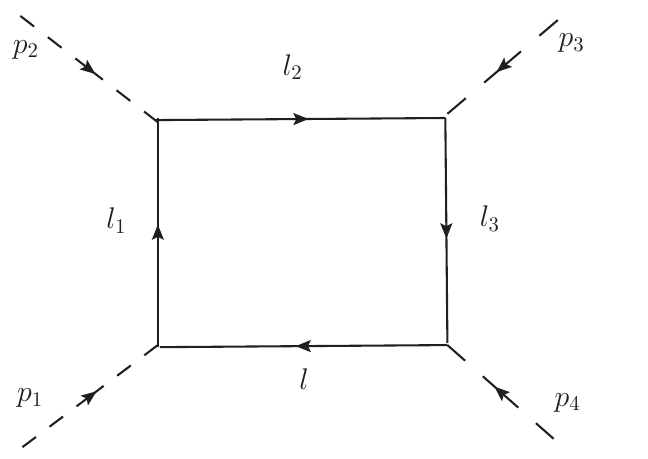}
\caption{Interaction of four scalars in the Yukawa theory of scalars and fermions.}
\label{fig:phi-4}
\end{center}
\end{figure}

We consider the scattering of four scalars via fermion loop diagram in Yukawa theory. 
The box amplitude of the fermion loop diagram, shown in Fig.~\ref{fig:phi-4}, is\footnote{Note 
that there are two more independent diagrams which contribute to the fermion loop $\phi^4$-amplitude 
in the Yukawa theory. We do not require them here.} 
\begin{eqnarray}
 {\cal M}_{1234}(\phi^4) &=& 
\int \frac{d^n l}{(2\pi)^n}\; \frac{{\rm tr}({\slashed l} {\slashed l_3} 
                                             {\slashed l_2} {\slashed l_1} )}{l^2\; l_1^2\; l_2^2\; l_3^2} 
\nn \\
&=&\int \frac{d^n l}{(2\pi)^n}\; \frac{4(l.l_3\;l_2.l_1 - l.l_2\;l_3.l_1 + l.l_1\;l_2.l_3)}
{l^2\; l_1^2\; l_2^2\; l_3^2}.
\end{eqnarray}
Rearranging the terms in the numerator, we can write
\begin{eqnarray}
 {\cal M}_{1234}(\phi^4) &=& (p_1^2 p_3^2 + p_2^2 p_4^2 - st)\; D_0 + (s - p_1^2 - p_2^2)\; C_0(3) \nn \\
                         && + (t - p_1^2 - p_4^2)\; C_0(2) + (s - p_3^2 - p_4^2)\; C_0(1) \nn \\
                         && + (t - p_2^2 - p_3^2)\; C_0(0) + 2\; B_0(0,2) + 2\; B_0(1,3),
\end{eqnarray}
where $s=(p_1+p_2)^2$ and $t=(p_2+p_3)^2$. The $D_0$ is the box scalar integral, 
\begin{equation}
 D_0 = \int \frac{d^n l}{(2\pi)^n}\; \frac{1}{d_0\; d_1\; d_2\; d_3}, 
\end{equation}
where $d_0=l^2$ and $d_i = l_i^2$.
$C_0(i)$ and $B_0(i,j)$ are reduced triangles  and bubbles,
written in the missing propagator notation. Since the bubble scalars
are IR finite, the IR finiteness of the above fermion loop amplitude implies                                                                                                 
\begin{eqnarray}\label{eq:IR-D0toC0}
 D_0|_{IR} &=& \frac{1}{(st - p_1^2 p_3^2 - p_2^2 p_4^2)}\Big[ (t - p_2^2 - p_3^2)\; C_0(0)|_{IR} +
                                                             (s - p_3^2 - p_4^2)\; C_0(1)|_{IR} \nn \\
&     & + (t - p_1^2 - p_4^2)\; C_0(2)|_{IR} + (s - p_1^2 - p_2^2)\; C_0(3)|_{IR} \Big],
\end{eqnarray}
which is the desired result. It can be compared with the identity, derived in \cite{Bern:1993kr} 
\begin{eqnarray}
 D_0^n = \frac{1}{2}\left( \sum_{i=0}^3 \alpha_i C_0^n(i) + (3-n)\beta D_0^{n+2}\right),
\end{eqnarray}
where $\alpha_i = \sum_{i=0}^3 (Y^{-1})_{i+1\;j+1}$ and $\beta = \sum_{i=0}^3 \alpha_i$. $Y_{ij}$
is the modified Cayley matrix of the box integral. Since the six-dimensional box integral is finite,
the IR structure of box integrals is completely decided by that of triangle integrals in 4 dimensions.
The identity in Eq.~\ref{eq:IR-D0toC0}, can now be used to obtain the IR singular terms in some of the 
complicated box scalar integrals. These are listed in table 1, and can be cross-checked with Ref.~\cite{Ellis:2007qk}. \\

\begin{table}\label{tab:d0_ir}

\begin{center}
\begin{tabular}{|l|c|l|}
 \hline
 {\bf Case} & {\bf $A$} & {\bf $B$} \\
 \hline
 $p_1^2 = p_2^2 = p_3^2 = p_4^2 = 0 $ & $\frac{4}{st}$ & $\frac{2}{st} \left[- {\rm ln}(-s) - {\rm ln}(-t)\right]$ \T\B \\
 \hline
 $p_1^2 = p_2^2 = p_3^2 = 0; p_4^2 \ne 0 $ & $\frac{2}{st}$ & $\frac{2}{st} \left[{\rm ln}(-p_4^2) - {\rm ln}(-s) - {\rm ln}(-t)\right]$ \T\B \\
 \hline
  $p_1^2 = p_2^2 = 0; p_3^2 \ne 0;  p_4^2 \ne 0 $ & $\frac{1}{st}$ & $\frac{1}{st} \left[{\rm ln}(-p_3^2) 
  +{\rm ln}(-p_4^2) -{\rm ln}(-s) - 2\; {\rm ln}(-t)\right]$ \T\B \\
 \hline
   $p_1^2 = p_3^2 = 0; p_2^2 \ne 0;  p_4^2 \ne 0 $ & 0 & $\frac{2}{(st-p_2^2p_4^2)} \left[ {\rm ln}(-p_2^2) + {\rm ln}(-p_4^2) 
   - {\rm ln}(-s) - {\rm ln}(-t)\right]$ \T\B \\
 \hline
   $p_1^2 = 0; p_2^2 \ne 0; p_3^2 \ne 0;  p_4^2 \ne 0 $ & 0 & $\frac{1}{(st-p_2^2p_4^2)} \left[ {\rm ln}(-p_2^2) + {\rm ln}(-p_4^2) 
   - {\rm ln}(-s) - {\rm ln}(-t)\right]$ \T\B \\
 \hline
\end{tabular}
 \end{center}
 \caption{IR singular pieces of box scalar integrals for various cases of external vertualities. All 
          the internal lines are massless. The IR singular structures $A$ and $B$ are explained in the 
          introduction.}
\end{table}


\section{Conclusion}
In the above, we have shown that the most general possible fermion loop diagram is IR finite. We have seen that the soft finiteness of 
fermion loop diagrams follows from simple power counting in vanishing loop momentum while their collinear finiteness results, utilizing 
various properties of massless external particles attached to the loop. We have verified this fact in many triangle, box and pentagon
type fermion loop diagrams.
We have further applied this result to obtain the IR singular structure of box
scalar integrals in terms of those of reduced triangle scalar integrals.
We would also like to emphasize the importance of this result 
from the point of view of making numerical checks on the calculation of complicated
fermion loop diagrams.

\section*{Acknowledgment}

The author would like to thank Pankaj Agrawal for his valuable comments on generalizing the result.


\begin{thebibliography}{20}

  
\bibitem{Kinoshita:1962ur} 
  T.~Kinoshita,
  J.\ Math.\ Phys.\  {\bf 3}, 650 (1962).

\bibitem{Collines89}
  J. C. Collines, in {\it Perturbative Quantum Chromodynamics}, ed. A. H. Muller (World Scientific, Singapore, 1989). 

\bibitem{Ellis:2007qk} 
  R.~K.~Ellis and G.~Zanderighi,
  JHEP {\bf 0802}, 002 (2008)
  [arXiv:0712.1851 [hep-ph]].


\bibitem{Dittmaier:2003bc} 
  S.~Dittmaier,
  Nucl.\ Phys.\ B {\bf 675}, 447 (2003)
  [hep-ph/0308246].

\bibitem{Shivaji:2010re} 
  A.~Shivaji,
  arXiv:1008.4375 [hep-ph].
  
\bibitem{Nagy:2003qn} 
  Z.~Nagy and D.~E.~Soper,
  JHEP {\bf 0309}, 055 (2003)
  [hep-ph/0308127].

\bibitem{Holstein06}
  B. R. Holstein,
  Am. J. Phys. {\bf 74}, 1002 (2006).

\bibitem{Bern:1993kr} 
  Z.~Bern, L.~J.~Dixon and D.~A.~Kosower,
  Nucl.\ Phys.\ B {\bf 412}, 751 (1994)
  [hep-ph/9306240].


\end{thebibliography}
\end{document}